**Physics-Trained Neural Network as Inverse Problem Solver for Potential Fields: An Example of Downward Continuation between Arbitrary Surfaces**

Jing Sun[1*], Lu Li[2], Liang Zhang[3]

1. Faculty of Electrical Engineering, Mathematics, and Computer Science, Delft University of Technology, The Netherlands (Jing.Sun@tudelft.nl)
2. Commonwealth Scientific and Industrial Research Organization (CSIRO), Australia
3. Key Laboratory of Advanced Manufacturing Technology of the Ministry of Education, Guizhou University, China

**Summary**

Downward continuation is a critical task in potential field processing, including gravity and magnetic fields, which aims to transfer data from one observation surface to another that is closer to the source of the field. Its effectiveness directly impacts the success of detecting and highlighting subsurface anomalous sources. We treat downward continuation as an inverse problem that relies on solving a forward problem defined by the formula for upward continuation, and we propose a new physics-trained deep neural network (DNN)-based solution for this task. We hard-code the upward continuation process into the DNN's learning framework, where the DNN itself learns to act as the inverse problem solver and can perform downward continuation without ever being shown any ground truth data. We test the proposed method on both synthetic magnetic data and real-world magnetic data from West Antarctica. The preliminary results demonstrate its effectiveness through comparison with selected benchmarks, opening future avenues for the combined use of DNNs and established geophysical theories to address broader potential field inverse problems, such as density and geometry modelling.

**Introduction**

Downward continuation of potential field, including gravity or magnetic field, refers to transferring the data from one observation surface to a lower surface that is closer to the source of the field. The goal is to enhance the resolution of the continued field and amplify the shallow geological signals. Airborne surveys are typically flown at uneven heights, making continuation from these surfaces a common requirement. Downward continuation is a critical task in the processing of potential field data, impacting the success of various downstream analyses, such as revealing the density structure and boundaries of anomalous bodies, especially for detecting and highlighting shallow anomalous sources.

Many methods have been developed for the task of downward continuation (e.g. Pilkington and Boulanger, 2017 and reference therein). Based on the numerical scheme, they can be classified into two categories: frequency-domain (e.g. Cooper, 2004) and space domain approaches (e.g. Guo and Tao, 2020). Frequency-domain approaches are based on the fast Fourier transform, which amplify short wavelength information. However, it causes downward continuation sensitive to noise level and highly unstable (Cooper, 2004).

In contrast, potential field continuation can be solved directly in the spatial domain. The upward continuation is equivalent to a matrix multiplication problem $KU_0 = U$, where $U_0$ is the potential field observed at the lower surface and $K$ is the kernel matrix. Note that in the context of this paper, "kernel" always refers to the geophysical concept, not the term used in deep learning (DL) to describe convolutional operations. It is essentially a matrix that records the space-domain difference between two surfaces. Conversely, downward continuation can be seen as an inverse problem derived from upward continuation, where $U$ is the observed field at the higher surface and $U_0$ is the unknown field, assumed observed at the lower surface.

In recent years, numerous DL applications have emerged in geophysics, with early efforts primarily focused on supervised learning (SL). However, a key challenge in geophysical tasks, including the downward continuation of potential fields, is the absence of ground truth data in real-world problem-solving scenarios. This underscores the need for developing new learning strategies that do not depend on labelled training pairs. Besides, well-established geophysical theories are often overlooked in data-driven SL approaches, despite their potential to offer valid guidance to the deep neural networks (DNNs)' learning process. Therefore, to address the unique challenges posed by geophysical problems, it is worthwhile to explore physics or theory-based DL approaches.

In this work, we propose a physics-driven self-supervised DL solution for potential field downward continuation in the spatial domain. The deterministic upward continuation process is hard-coded into the DNN's learning process, steering the DNN to learn to function as an inverse problem solver, that is, to downward continue the potential field, without ever being shown the ground truth data. We test our proposed method using both synthetic and real-world data. The real-work application uses aeromagnetic data from Thwaites Glacier in West Antarctica (Jordan et al., 2023). Results demonstrate our proposed method could downward continuate aeromagnetic data between arbitrary surfaces and effectively highlighting local subglacial geological features.

**Upward Continuation**

Downward continuation is regarded as an inverse problem, for which the forward problem is upward continuation. Guo and Tao (2020) proposed a kernel function-based solution for upward continuation in the space domain without limitation on the continuation distance, which serves as the foundation within the framework of our proposed method. The formula of upward continuation from Blakely (1996) can be written as,

$$U(x, y, z) = \frac{\Delta z(x, y)}{2\pi} \int_{-\infty}^{\infty} \int_{-\infty}^{\infty} \frac{U(\alpha, \beta, z_0)}{R^3} d\alpha d\beta \tag{1}$$

where $U(\alpha, \beta, z_0)$ is the observed field on a horizontal plane $S$, and $U(x, y, z)$ is the upward continued field on an irregular surface $Q$. $\Delta z = z(x, y) - z_0 \geq 0$ is a function of $(x, y)$ rather than a constant value as determined via integration. $R = [(x - \alpha)^2 + (y - \beta)^2 + \Delta z^2]^{1/2}$ is the distance between any upward continued point $P$ on surface $Q$ and any observed point $M$ on horizontal plane $S$ (Figure 1a).

Guo and Tao (2020) reduce Eq. 1 into two entities with a finite integration which can be expressed as,

$$U(x, y, z) = \frac{\Delta z(x, y)}{2\pi} \int_{\alpha_1}^{\alpha_2} \int_{\beta_1}^{\beta_2} \frac{U(\alpha, \beta, z_0)}{R^3} d\alpha d\beta + \sum_{i=1}^{4} O(S_i) \tag{2}$$

where $[\alpha_1, \alpha_2]$ and $[\beta_1, \beta_2]$ represent the dimensions of observation in directions $\alpha$ and $\beta$, respectively. $O(S_i)$ ($i = 1,2,3,4$) is the integration for the outside infinite regions (Figure 1b).

Eq. 2 can be reduced to the sum of piecewise integrals and discretized by the midpoint quadrature rule (Figure 1b) as,

$$U_{m,n} = \frac{1}{2\pi} \sum_{i=1}^{M} \sum_{j=1}^{N} P_{m,n;ij} U_{ij} \tag{3}$$

where $U_{m,n} = U(x_m, y_n, z)$ is the continued field which is discretized into a similar grid of observational field $U_{ij} = U(\alpha_i, \beta_j, z_0)$. $U_{ij}$ is assumed as a constant in subdomain of integration (blue translucent rectangle in Figure 1b). The piecewise integral $P_{m,n;ij}$ can be written as,

$$P_{m,n;ij} = \Delta z(x_m, y_n) \int_{\alpha_i - \Delta x/2}^{\alpha_i + \Delta x/2} \int_{\beta_j - \Delta y/2}^{\beta_j + \Delta y/2} \frac{d\alpha d\beta}{R^3} \tag{4}$$

where $\alpha_i = \alpha_1 + i \cdot \Delta x$ and $\beta_j = \beta_1 + j \cdot \Delta y$ represent the central points of subdomain of piecewise integration. Combining Eq. 3 and 4, the upward continuation integration can be converted into a matrix form, $U = KU_0$, where $U$ and $U_0$ represent vector of discretized continued and observed field, respectively. $K$ is the upward continuation kernel matrix with dimensions $(M \times N, M \times N)$ where

$$k_{m,n;ij} = \frac{1}{2\pi} P_{m,n;ij} . \tag{5}$$

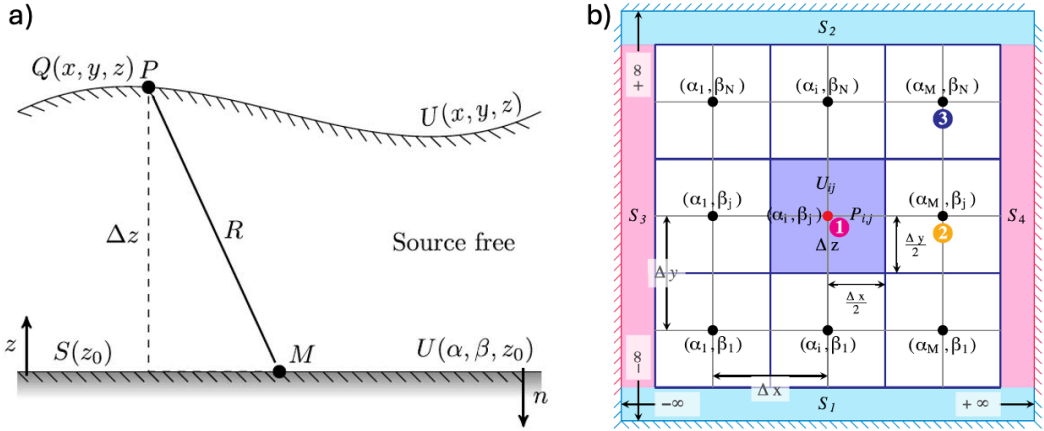

*Figure 1. (a) Geometry for upward continuation from a horizontal plan to an arbitrary surface. (b) Schematic diagram of integral and discretization schemes (adapted from Guo and Tao, 2020).*

**The Proposed Physics-Driven Self-Supervised Deep Learning Method**

Figure 2 illustrates the framework of the proposed method. The DNN employed in this paper consists of residual blocks (He et al., 2016), which were initially introduced as part of ResNet. While the current framework leverages a specific architecture based on residual connections, exploring different network architectures is worthwhile for further performance enhancement.

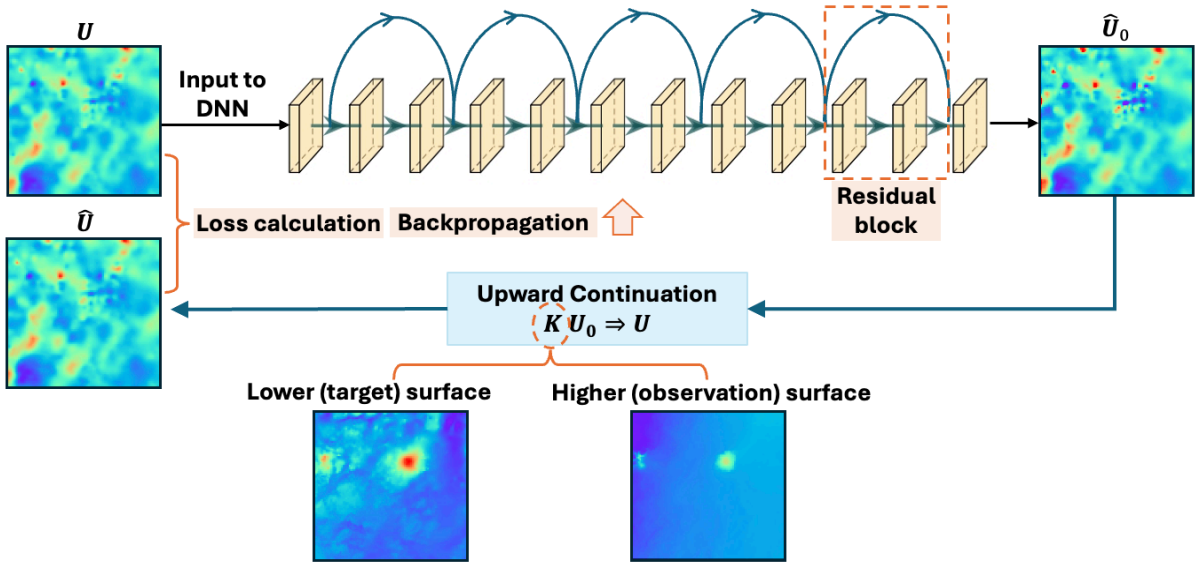

*Figure 2. The framework of the physics-driven DL method with a visualization of the employed residual block-based DNN model. $U$ is the observed potential field. $\hat{U}_0$ is the DNN's output, that is, the downward continued field obtained by taking $U$ as the input. $K$ is the geophysical kernel matrix derived from the height maps of the observation and the target surfaces. $\hat{U}$ is the upward continued field based on the DNN's output $\hat{U}_0$.*

The DNN processes the observed potential field $U$ as input, and its output $\hat{U}_0$ (the downward continued field), rather than being compared with the ground truth, is passed through the upward continuation function, where the required kernel matrix $K$ is derived from the height maps of the observation and the target surfaces. The height maps of both surfaces are recorded during acquisition. Afterward, the upward continued field based on the DNN's output is used for loss calculation with the original input to the DNN, upon which the backpropagation is conducted.

**Synthetic Data Example**

To demonstrate the effectiveness of our proposed method, we first test it on synthetic magnetic data generated by a 3D intrusion model using Noddy (Jessell and Valenta, 1996). The model domain includes a 2×2×2 km volume with 80 m voxels. We calculate the magnetic response in 100 m above the flat top surface and an arbitrary undulated surface (Figures 3a). Our goal is to downward continue the magnetic field (Figure 3b) to 100 m. Figure 3c shows the upward-continued field based on the DNN's output; the loss is computed by comparing it with the DNN's input (Figure 3b). Their final difference after the DNN's learning process is shown in Figure 3d.

For evaluation, we display the magnetic field simulated at the target surface of 100 m in Figure 3e and the DNN's output in Figure 3f, as well as their misfit in Figure 3g. The downward continuation result highlights the distribution of individual magnetic intrusions, in contrast to the continuous linear magnetic high observed at higher elevations. We observe that the misfit maps in Figures 3d and 3g exhibit spatial similarities with the input magnetic data, which is largely attributed to the resolution of the observational data and the continuation height (Guo and Tao, 2020).

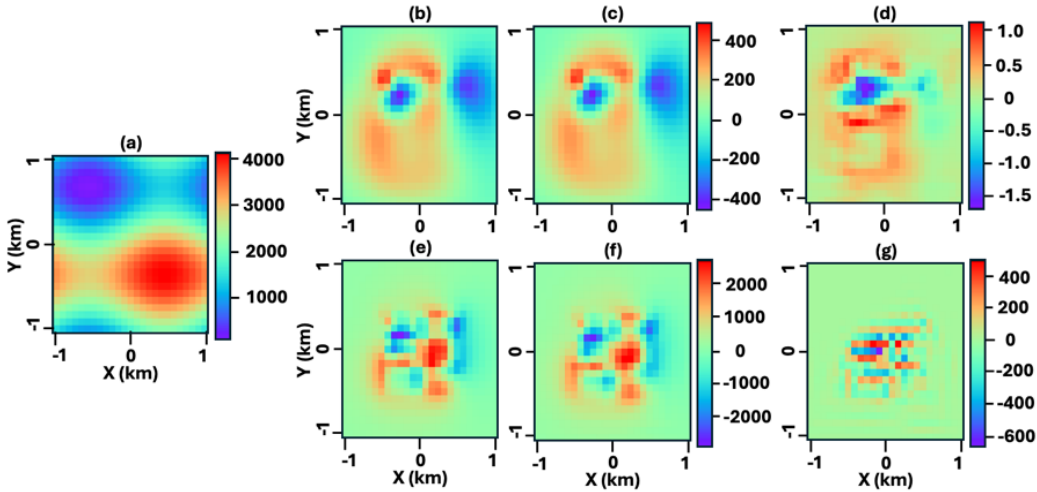

*Figure 3. Synthetic data examples: (a) height map of the observation surface; (b) observed magnetic field, serving as input to the employed DNN; (c) upward-continued field based on the DNN's output, used for loss computation; (d) difference between (b) and (c); (e) magnetic field simulated at the target surface; (f) DNN's output; (g) difference between (e) and (f).*

**Field Data Example**

We also test the proposed method on real-world magnetic field from Thwaites Glacier, West Antarctica (Jordan et al., 2023). The airborne magnetic data were collected between 2004 and 2020 with varying line spacings and flight heights. These lines are leveled and interpolated onto two 1×1 km cellwise grids: one with each line upward and downward continued at an elevation of 2500 m above the bedrock topography (Figures 4a) and the other continued to 500 m above the ice surface (Figure 4b). To evaluate the proposed method for downward continuation between arbitrary surfaces, we use the magnetic field at 2500 m above the bedrock topography as the observed (to-be-downward-continued) field (Figure 4c), and the magnetic field at 500 m above the ice surface serves as the benchmark result (Figure 4d).

The downward continuation result from the proposed method is shown in Figure 4e. We compare the upward-continued fields (Figures 4f and 4g) based on respectively the benchmark's and the DNN's outputs with the observed magnetic field (Figure 4c). The corresponding differences are shown in Figures 4h and 4i, respectively. It is clear that the upward-continued field based on the DNN's output is much closer to the observed magnetic field, as Figure 4i illustrates a relatively smaller difference. The larger difference in Figure 4h can be attributed to the use of different upward continuation techniques (1D line-by-line vs. 2D kernel). Figure 4j shows the difference between the downward continuation results obtained from the benchmark and the proposed method. Compared to the benchmark's result (Figure 4d), the proposed method produces a similar overall result but with refined magnetic textures and amplitudes. For instance, the DNN output reveals that Mt. Takahe (a snow-covered shield volcano) exhibits a high-amplitude magnetic low, potentially indicating stronger magnetic remanence. Additionally, the linear magnetic high, which indicates mafic intrusion, shows improved local refinement and sharpness in the result.

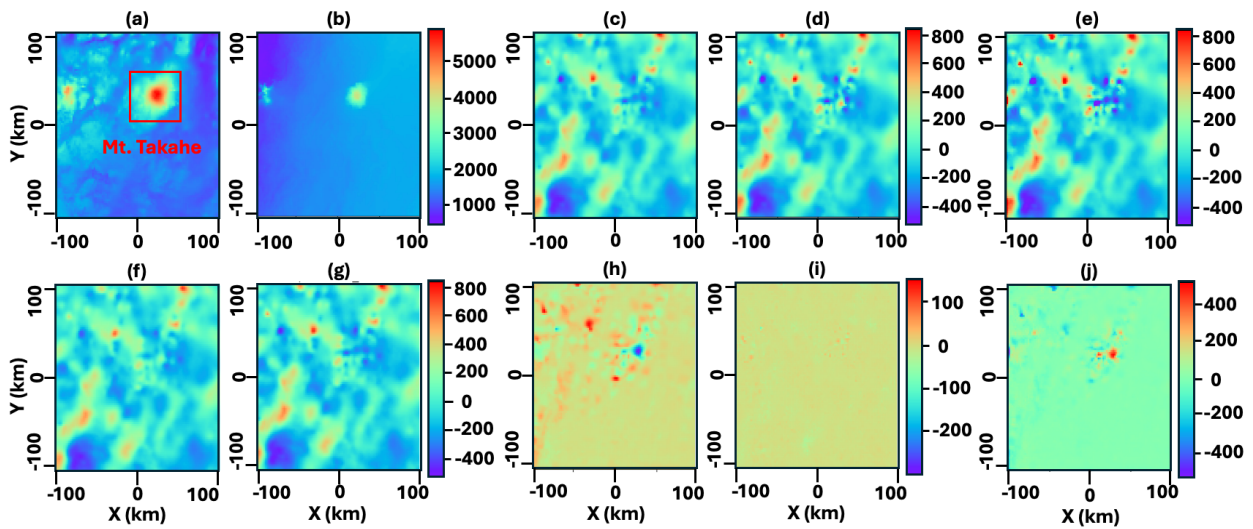

*Figure 4.* Field data examples from West Antarctica: (a) and (b) are height maps of the observation surface and the targeted surface, respectively; (c) is the magnetic field observation at surface (a); (d) is the benchmark result and (e) is the proposed method's result (i.e., DNN's output) on downward continuation; (f) and (g) are upward-continued fields respectively based on (d) and (e); (h) and (i) are differences respectively between (f) and (g) with (c); (j) is the difference between (d) and (e).

**Conclusions**

In this work, we treat downward continuation as an inverse problem, for which the forward problem is upward continuation. We employ a DNN as the inverse problem solver, whose learning process is driven by the embedded physical function of upward continuation, rather than a large amount of 'labelled' training pairs, thus eliminating dependency on the existence of ground truth data. We demonstrate its effectiveness in downward continuation of magnetic fields between arbitrary surfaces using synthetic and real-world data. This approach opens future avenues for the combined use of DNNs and established geophysical theories to address broader potential field inverse problems, such as density and geometry modelling.